\documentclass[aps,prl,twocolumn,nofootinbib,showpacs,preprintnumbers,notitlepage,groupedaddress, nobalancelastpage]{revtex4-2}

\usepackage{amsmath,amsfonts,amssymb}
\usepackage[utf8]{inputenc}
\usepackage{braket}
\usepackage{graphicx}
\usepackage{bm}
\usepackage{placeins}
\usepackage{xfrac}
\usepackage[colorlinks=true, breaklinks=true, linkcolor=red, citecolor=blue, urlcolor=blue]{hyperref} 
\usepackage[capitalize]{cleveref}
\usepackage{changes}
\definechangesauthor[color=blue]{tom}
\definechangesauthor[color=green]{lv}
\allowdisplaybreaks
\usepackage{kantlipsum}

\newcommand{\x}{\ensuremath{\mathbf{x}}}

\definecolor{tensorblue}{rgb}{0.8,0.8,1}
\definecolor{lighttensorblue}{rgb}{0.9,0.9,1}
\definecolor{tensorred}{rgb}{1,0.5,0.5}
\definecolor{tensorpurp}{rgb}{1,0.5,1}

\tikzset{ten/.style={fill=tensorblue}}
\tikzset{lightten/.style={fill=lighttensorblue}}
\tikzset{tenred/.style={fill=tensorred}}
\tikzset{tengreen/.style={fill=green!50!black!20}}
\tikzset{tenpurp/.style={fill=tensorpurp}}
\tikzset{tengrey/.style={fill=black!20}}
\tikzset{tenorange/.style={fill=orange!30}}
\tikzset{tenwhite/.style={fill=white}}
\tikzset{u/.style={fill=blue!20,draw=black}}
\tikzset{w/.style={fill=green!50!black!80,draw=black}}

\tikzstyle myBG=[line width=3pt,opacity=1.0]

\usepackage{ bbold }

\let\revappendix\appendix

\usepackage{tikz}
\usetikzlibrary{arrows,positioning} 
\usetikzlibrary{arrows.meta,decorations.markings, snakes}
\usetikzlibrary{math} 
\usepackage{pgf}
\tikzset{
    >=stealth',
    pil/.style={
           ->,
           thick,
           shorten <=2pt,
           shorten >=2pt,}
}

\newcommand\minus{%
  \setbox0=\hbox{-}%
  \vcenter{%
    \hrule width\wd0 height \the\fontdimen8\textfont3%
  }%
}

\tikzset{->-/.style={decoration={
  markings,
  mark=at position #1 with {\arrow{>}}},postaction={decorate}}}
  
\tikzset{-<-/.style={decoration={
  markings,
  mark=at position #1 with {\arrow{<}}},postaction={decorate}}}

\newcommand{\scale}[0]{0.45}

\newcommand{\diagramone}[0]{\begin{array}{c}
\begin{tikzpicture}[scale=\scale]
\foreach \x in {0,2,...,6}{
\draw[shift={(\x,0)}] (0,0) -- (0,6);
\draw[shift={(0,\x)}] (0,0)--(6,0);
};
\foreach \x in {0,2,...,6}{
    \foreach \y in {0,2,...,6}{
        \draw (\x+1/4,\y+1/4)--(\x+4/4,\y+4/4);
    	\filldraw[tenwhite] (\x+4/4,\y+4/4) circle (0.40);
    	\node at(\x+4/4,\y+4/4) {$x$};
    	\filldraw[ten,shift={(\x,\y)}]  (-1/2,-1/2)--(1/2,-1/2)--(1/2,1/2)--(-1/2,1/2)--(-1/2,-1/2);
	};
};
\end{tikzpicture}
\end{array}}

\newcommand{\diagramtwo}[0]{\begin{array}{c}
    \begin{tikzpicture}[scale=\scale]
    \draw (-5/2,0)--(5/2,0);
    \draw (0,5/2)--(0,-5/2);
    \draw (0,0)--(3/4,3/4);
    \filldraw[ten] (-1/2,-1/2)--(1/2,-1/2)--(1/2,1/2)--(-1/2,1/2)--(-1/2,-1/2);
    \filldraw[tenorange] (1,-1/2)--(2,-1/2)--(2,1/2)--(1,1/2)--(1,-1/2);
    \filldraw[tenorange] (-1/2,1)--(1/2,1)--(1/2,2)--(-1/2,2)--(-1/2,1);
    \filldraw[tenorange] (-1,-1/2)--(-2,-1/2)--(-2,1/2)--(-1,1/2)--(-1,-1/2);
    \filldraw[tenorange] (-1/2,-1)--(1/2,-1)--(1/2,-2)--(-1/2,-2)--(-1/2,-1);
    \node at(0,0) {$A$};
    \node at(3/2,0) {$B$};
    \node at(0,3/2) {$B$};
    \node at(-3/2,0) {$B$};
    \node at(0,-3/2) {$B$};
    \end{tikzpicture}
    \end{array}}
    
\newcommand{\diagramthree}[0]{\begin{array}{c}
    \begin{tikzpicture}[scale=\scale]
    \draw (-1,0)--(1,0);
    \draw (0,1)--(0,-1);
    \draw (0,0)--(3/4,3/4);
    \filldraw[ten] (-1/2,-1/2)--(1/2,-1/2)--(1/2,1/2)--(-1/2,1/2)--(-1/2,-1/2);
    \node at(0,0) {$\tilde{A}$};
    \end{tikzpicture}
    \end{array}}
    
\newcommand{\diagramfour}[0]{\begin{array}{c}
  \begin{tikzpicture}[scale=\scale]
  \draw (-1,0)--(1,0);
  \draw (0,1)--(0,-1);
  \draw (0,0)--(3/4,3/4);
  \node at(-1.2,0) {$0$};\node at(0,-1.4) {$0$};\node at(0,1.4) {$0$};
  \node at(1.2,0) {$0$};
  \filldraw[ten] (-1/2,-1/2)--(1/2,-1/2)--(1/2,1/2)--(-1/2,1/2)--(-1/2,-1/2);
  \node at(1,1) {$0$};
  \end{tikzpicture}
  \end{array}}
  
\newcommand{\diagramfive}[0]{\begin{array}{c}
  \begin{tikzpicture}[scale=\scale]
  \draw (-1,0)--(1,0);
  \draw (0,1)--(0,-1);
  \draw (0,0)--(3/4,3/4);
  \node at(-1.2,0) {$1$};\node at(0,-1.4) {$0$};\node at(0,1.4) {$0$};
  \node at(1.2,0) {$1$};
  \filldraw[ten] (-1/2,-1/2)--(1/2,-1/2)--(1/2,1/2)--(-1/2,1/2)--(-1/2,-1/2);
  \node at(1,1) {$1$};
  \end{tikzpicture}
  \end{array}}
  
\newcommand{\diagramsix}[0]{\begin{array}{c}
  \begin{tikzpicture}[scale=\scale]
  \draw (-1,0)--(1,0);
  \draw (0,1)--(0,-1);
  \draw (0,0)--(3/4,3/4);
  \node at(-1.2,0) {$0$};\node at(0,-1.4) {$1$};\node at(0,1.4) {$1$};
  \node at(1.2,0) {$0$};
  \filldraw[ten] (-1/2,-1/2)--(1/2,-1/2)--(1/2,1/2)--(-1/2,1/2)--(-1/2,-1/2);
  \node at(1,1) {$1$};
  \end{tikzpicture}
  \end{array}}

\begin{document}
\title{Generative modeling with projected entangled-pair states}

\author{Tom Vieijra}
\author{Laurens Vanderstraeten}
\author{Frank Verstraete}

\affiliation{Department of Physics and Astronomy, Ghent University, B-9000 Ghent, Belgium}

\begin{abstract}
We argue and demonstrate that projected entangled-pair states (PEPS) outperform matrix product states significantly for the task of generative modeling of datasets with an intrinsic two-dimensional structure such as images. Our approach builds on a recently introduced algorithm for sampling PEPS, which allows for the efficient optimization and sampling of the distributions.
\end{abstract}

\maketitle

\par\noindent\emph{\textbf{Introduction---}} %
%
Driven by advances in computational hardware such as graphical processing units and research in optimization and expressibility, recent progress in machine learning and artificial intelligence has allowed analyzing and modeling large amounts of high-dimensional data. These models and methods are now extensively used in numerous applications, both in industry and in science \cite{RevModPhys.91.045002}. A particularly general approach to capturing the features of a given dataset is modeling its underlying probability distribution. This subfield is called generative modeling, and allows to apply statistical techniques on the distribution, such as drawing samples from it, and calculating marginal and conditional distribution of the variables. Techniques from many-body physics have always played a major role in the development of generative machine learning \cite{Hopfield2554, ackley1985learning}. This can be traced back to the parallels between the respective problems one has to deal with in both fields. For example, classical many-body physics concerns modeling the partition function of a collection of many degrees of freedom, and subsequently evaluating expectation values from it.
\par An explicit example of this import of ideas from many-body physics into machine-learning applications is the use of energy-based models \cite{lecun2006tutorial, Du19}. Here the probability distribution is modeled as a Boltzmann distribution, where the energy can be interpreted as a sum of interactions between the variables of the distribution. These interactions contain variational parameters that can be optimized such that the model approaches the data distribution, a problem known as the inverse Ising problem \cite{Aurell2012}. A particular example of such energy-based models is the restricted Boltzmann machine (RBM). It has been proven that RBMs can approximate any distribution to arbitrary precision, but this does not imply that they do so efficiently~\cite{LeRouxRepresentational08, LeRouxDeep10}. In particular, the RBM does not take into account the symmetries and locality inherent in the dataset. It has been pointed out that efficient descriptions of classical data can be devised with the help of the structure of mutual information between subsets of variables~\cite{cheng2018, lu2021tensor, martyn2020entanglement}, which provides an inductive bias for models that can efficiently capture the data.
\par In the field of quantum many-body physics, the locality of interactions has profound consequences for the efficient representability of the relevant quantum states. In particular, it was realized that the ground states of local Hamiltonians exhibit an area law for the entanglement entropy, and can therefore be approximated efficiently as tensor network states~ \cite{Verstraete2006}. On the other hand, a Boltzmann distribution with local interactions can be naturally formulated as a tensor network~\cite{LiBoltzmann21,ClarkUnifying18}. Tensor networks have been used for machine-learning tasks in the context of classification \cite{Stoudenmire2016, Stoudenmire_2018, glasser2020probabilistic, Cheng2021}, as a means of compressing neural networks \cite{WU2020309} and as a model for the underlying distribution of datasets \cite{Glasser19, Stokes19, Han2018, Cheng2019, liu2021tensor}. The particular structure of tensor networks also allows to investigate properties of the datasets and how the network models it, something which is particularly attractive compared to black-box models \cite{Bengua15, Bengua16, Bradley_2020}. A crucial aspect in applying tensor networks for modeling a given dataset, is that the network should mimic the local structure of the data in an efficient way. For datasets with an intrinsic two-dimensional structure such as images, the use of projected entangled-pair states (PEPS) seems to be the natural choice. A straightforward application of PEPS to the generative modeling of images has been prohibited by a lack of efficient algorithms and, therefore, previous works have taken recourse to matrix product states \cite{Han2018} and tree tensor networks \cite{Cheng2019}. The application of PEPS has until now been restricted to classification problems \cite{Cheng2021}, but are lacking in generative modeling. 
\par In this work, we fill this gap and apply PEPS to generative modeling of datasets that have an intrinsic two-dimensional structure. We start by reformulating local energy-based models in terms of PEPS wavefunctions, after which we can generalize this approach in a natural way. We explain how a direct sampling algorithm \cite{Vieijra2021} can be used for optimizing a general PEPS ansatz with respect to the log-likelihood for a given dataset. Our first example is the bars and stripes dataset, for which an exact PEPS can be written down, and afterwards we apply our scheme to the MNIST set of handwritten numbers.

\par\noindent\emph{\textbf{Generative modeling with PEPS---}} %
%
Consider a labeled dataset $X$, consisting of tuples $(\mathbf{x}, \mathbf{y}) \in X$, where $\mathbf{x}$ is the data point and $\mathbf{y}$ is the associated label. Every data point $\mathbf{x}$ is a vector in a $N$-dimensional space. The labels are encoded as unit vectors in a $n_c$-dimensional space where $n_c$ is the number of distinct labels. This dataset can be thought of as originating from sampling the distribution $P(\mathbf{x},\mathbf{y})$, i.e. the joint probability distribution of the data points and the labels. The aim of machine learning methods is modeling this distribution, or its derived marginal or conditional distributions. For example, in discriminative machine learning, one aims to find a model for the distribution $P(\mathbf{y}|\mathbf{x})$. In generative machine learning, one aims to approximate $P(\mathbf{x})$ or $P(\mathbf{x}|\mathbf{y})$.
\par As mentioned in the introduction, Boltzmann distributions in the form of energy-based models can be used to model such datasets. For local interactions, such a distribution can be naturally represented by a PEPS wavefunction.  For example, the Boltzmann distribution of an Ising model can be constructed by the tensor network
\begin{equation} \label{eq:single_layer}
\Psi(\mathbf{x}) = \diagramone,
\end{equation}
such that $P(\mathbf{x}) = \left| \Psi(\mathbf{x}) \right|^2$ is the Boltzmann weight. Indeed, by choosing five-leg tensors as
\begin{equation} \label{eq:absorb_int}
\diagramtwo = \diagramthree.
\end{equation}
with
\begin{equation}
A_{\alpha \beta \gamma \delta}^x = \begin{cases} 1, & \textrm{if}\ \alpha=\beta=\gamma=\delta=x \\ 0, & \textrm{otherwise}, \end{cases}
\end{equation}
and
\begin{equation}
B = \begin{bmatrix} e^{\beta/2} & e^{-\beta/2} \\ e^{-\beta/2} & e^{\beta/2} \end{bmatrix}^{\frac{1}{2}},
\end{equation}
the contraction, i.e. performing all tensor contractions defined by the edges between tensors in Eq.~\eqref{eq:single_layer}, yields the Boltzmann weight for configuration $\mathbf{x}$.  

\par The construction of a Boltzmann distribution with PEPS can be generalized by allowing complete freedom of all tensor elements. This corresponds to the formulation of variational quantum-mechanical states with PEPS. Performing the contraction yields a scalar value $\Psi(\mathbf{x})$, that is in general not real and positive. Because we aim to model a probability distribution $P(\mathbf{x})$, we require $P(\mathbf{x})\geq0$. To enforce this constraint we define $P(\mathbf{x}) = \left| \Psi(\mathbf{x}) \right|^2$, coinciding with the Born probability of states in quantum mechanics.
\par It is clear that the most efficient way of describing data with intrinsic two-dimensional correlations between the degrees of freedom is by a model that adheres to this geometric structure. An additional structural property of many real-world datasets is that the data is distributed into different disconnected modes. The tensor-network representation of a distribution that consists of a mixture of macroscopically different distributions can be written with tensors that are approximately block-diagonal (after fixing an appropriate gauge freedom). On the level of the states, this corresponds to the total wavefunction being a superposition of different wavefunctions with smaller entanglement (i.e., a cat state). Therefore, representing the total distribution as a sum of tensor-network states provides a significant compression in terms of number of parameters compared to the distribution described by a single tensor-network state.
\par For this paper, we will work directly with the decomposed picture and we propose the following distribution as our model:
\begin{equation}
  P(\mathbf{x}) = \frac{N_1}{N}P_1(\mathbf{x}) + ... + \frac{N_m}{N}P_{m}(\mathbf{x}).
  \label{eq:mixture}
\end{equation}
Here, we assume a sum over $m$ modes, where every mode $i$ has a weight $N_i/N$ determined by the fraction of data samples contained in mode $i$ in the training set. $P_i(\mathbf{x}) =\left| \Psi_i(\mathbf{x}) \right|^2 $ is the probability distribution defined by a PEPS tensor network.
\par To assign the data into modes, we propose two strategies. The first strategy is useful when one has access to a labeled dataset. One could then assign one mode to every label. Note that in this case, our model also allows to perform discriminative tasks, as one can infer the label of a given data point $\mathbf{x}$ by finding the mode that assigns the largest probability to the data point $\mathbf{x}$. The second strategy is assigning modes by specifically separating the dataset in disjunct subsets. This can be done e.g. by performing a dimensionality reduction such as UMAP \cite{2018arXivUMAP} and using the low-dimensional representation in a clustering algorithm such as DBSCAN \cite{Ester96adensity-based}. In this way, we partition the dataset into subsets by specifically searching for disjunct clusters that can be viewed as each originating from sampling a mode of the probability distribution. Note that this strategy is agnostic to the labeling, and hence unsupervised, while the first strategy can be seen as a form of supervised generative modeling.

\par\noindent\emph{\textbf{Optimization algorithm---}} %
%
Let us now explain how to optimize an individual PEPS tensor network by maximizing the log-likelihood for each mode of the training set. For the $i$-th mode, this cost function is defined as
\begin{equation} \label{eq:cost}
  \mathcal{L}_i = \sum_{\mathbf{x} \sim \mathrm{data}_i} \log (P_i(\mathbf{x})),
\end{equation}
where $\mathrm{data}_i$ is the distribution of data that is assigned to mode $i$. Here, we have defined \begin{equation}
    P_i(\mathbf{x}) = \frac{|\Psi_i(\x)|^2} {\sum_{\x'} |\Psi_i(\x')|^2 }
\end{equation}
where the sum runs over all possible configurations of $\mathbf{x}$. The numerator can be evaluated as the contraction of a single-layer tensor network, which we can do approximately with boundary-MPS methods \cite{Verstraete2004RenormalizationDimensions}, whereas the denominator represents the squared norm of the PEPS wavefunction. As introduced in Ref.~\onlinecite{Vieijra2021} we can compute the norm according to a direct-sampling algorithm. 
\par The optimization of the parameters in the PEPS tensors is performed by gradient-based methods, where the gradient of the cost function in Eq.~\eqref{eq:cost} to one of the real-valued tensors $A_j$ is
\begin{align}
  \frac{\partial \mathcal{L}_i}{\partial A_j} &= 2\sum_{\mathbf{x} \sim \mathrm{data}_i} \frac{1}{\Psi_i(\mathbf{x})} \frac{\partial \Psi_i(\mathbf{x})}{\partial A_j} - 2\sum_{\mathbf{x}} \frac{P_i(\mathbf{x})}{ \Psi_i(\mathbf{x})} \frac{\partial \Psi_i(\mathbf{x})}{\partial{A_j}} \nonumber \\
  &= 2\sum_{\mathbf{x} \sim \mathrm{data}_i} \frac{\partial \log \Psi_i(\mathbf{x})}{\partial A_j} - 2\sum_{\mathbf{x} \sim \mathrm{model}_i} \frac{\partial \log \Psi_i(\mathbf{x})}{\partial{A_j}}.
  \label{eq:gradient}
\end{align}
The first term of Eq.~\eqref{eq:gradient} is sometimes called the positive phase, while the second term is called the negative phase. The positive phase can again be computed by a single-layer contraction, where now the tensor $A_j$ is left out of the diagram consisting of tensors in Eq.~\eqref{eq:single_layer}, whereas the negative phase is again estimated by sampling. This implies that we use a stochastic optimization algorithm, which proves to be quite stable.
\par After having optimized the PEPS wavefunction for each mode, we can draw samples to generate new configurations. Here, again, it is important to draw independent samples from the distribution $P(\mathbf{x})$, which is made possible by the direct sampling algorithm of Ref.~\onlinecite{Vieijra2021}. Note that in generative machine learning, the distribution $P(\mathbf{x})$ often takes a form with many local maxima, separated by regions of low probability. Therefore, sampling algorithms such as Markov chains are infeasible due to autocorrelation effects, and a direct sampling of the distribution is preferred. 

\par\noindent\emph{\textbf{Bars and stripes---}} %
%
To show the importance of adhering to the geometric structure of data, we study the bars and stripes dataset. This is a toy dataset that consists of binary variables laid out in a two-dimensional grid. The data points are configurations such that for any given column or row all variables have the same binary value. When showing these data points in a two-dimensional grid by coloring the variables according to their binary value, e.g. black and white, the data points are configurations of either vertical bars or horizontal stripes. For a $4 \times 4$ grid, there are 30 unique configurations, each with probability $P(\mathbf{x}) = 1/30$, resulting in a log-likelihood of $\mathcal{L} = -3.4012$. 
\par The intrinsic two-dimensional structure of PEPS allows to efficiently describe states or distributions built from configurations with local constraints. This has been appreciated in a physical context, for example to describe dimer covering problems~\cite{Verstraete2006, Schuch2012, Vanderstraeten2018}. For these problems, the PEPS tensors can be constructed manually in such a way that the local constraints are satisfied and that the complete tensor network describes superpositions of configurations where the constraints are satisfied everywhere. The bars and stripes dataset fits in this class of locally-constrained models and we can write down a PEPS to capture the allowed configurations. Specifically, consider the PEPS with $D=2$ where all tensor elements are zero, except
\begin{equation} \label{eq:tensor}
   \diagramfour = \diagramfive = \diagramsix = 1.
\end{equation}
This tensor network has equal probabilities for all configurations in the dataset, except the one where all variables are equal to 1, that has a probability $4$ times higher than the other configurations. Configurations not in the dataset have zero probability. In total, this $D=2$ PEPS captures the complete data distribution, except one data point. Calculating the log-likelihood of this PEPS, we find $\mathcal{L} = -3.4503$, which is very close to the optimum. We have confirmed that starting from a random PEPS, we can converge to the same value for the log-likelihood by using the above optimization algorithm (not shown).
\par The bars and stripes dataset was also studied using an MPS-based model in Ref.~\cite{Han2018}. The MPS-based model is able to capture the complete dataset with a $D_{max}=16$ MPS. Note that the MPS in Ref.~\cite{Han2018} is equivalent to that obtained by hardcoding a configuration of a row in every virtual dimension of the MPS. This is a consequence of the fact that with MPS, one bond should propagate all information necessary between two rows of the configuration. In this way, the MPS acts as a memory for the specific configurations in the dataset. More generally, an MPS converges to a memorization of the dataset when the bond dimension approaches the number of training configurations. This property is potentially harmful for the generalization power of an MPS. Conversely, our PEPS construction takes advantage of the structural properties of the dataset, i.e. the correlation of values that are neighboring on the two-dimensional grid. Hence, this model forms a more natural representation of data with local correlations in two-dimensions, as the correlations the model can capture are treated equally in both directions. In addition, the PEPS parametrization is clearly extensive, in the sense that we can treat arbitrarily large grids without having to increase the bond dimension, in contrast to the MPS parametrization for which the bond dimension would scale exponentially with the linear dimension of the grid.

\par\noindent\emph{\textbf{MNIST dataset---}} %
%
\begin{figure}
\includegraphics{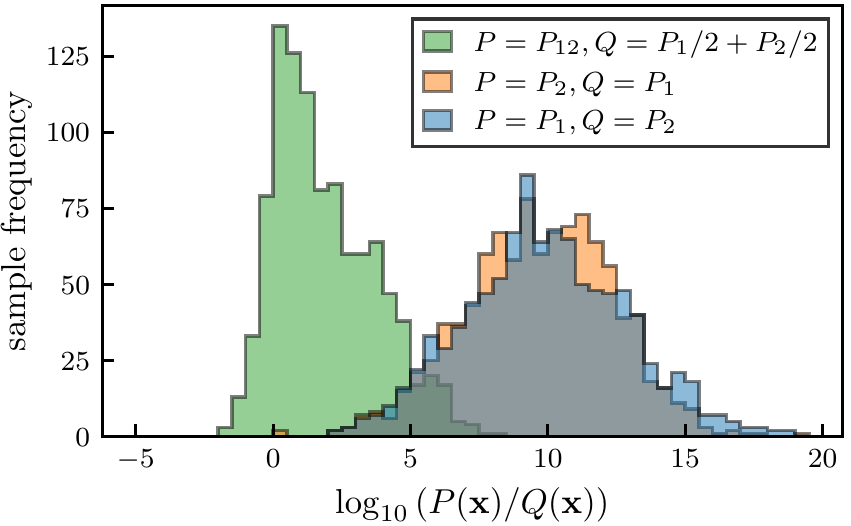}
\caption{Histogram of the difference in log-likelihoods under distributions $P$ and $Q$ for random samples . The samples are drawn according to the distribution $P$. Orange and blue compare likelihoods of samples under different modes. Green compares likelihoods between the model trained on both modes at the same time and the mixture of two models trained on the separate modes.}
 \label{fig:histogram}
\end{figure}
To benchmark the optimization of our model, we turn to the MNIST dataset that consists of greyscale images of handwritten digits. For details about the dataset, see Appendix.
First, we turn to a $8\times8$ downsizing of the MNIST data points to validate our assumption that a single PEPS modeling a multimodal distribution is equivalent to multiple PEPS modeling the modes separately. We optimize two $D=2$ PEPS with distribution $P_1$ and $P_2$ corresponding to two modes of the data. Next, we also model a $D=4$ PEPS with distribution $P_{12}$ on both modes at the same time. We can view the mixture distribution $P_1/2 + P_2/2$ of the $D=2$ PEPS distributions as that of a $D=4$ PEPS with block-diagonal tensors. We argued that the distribution of this PEPS should be close to the directly optimized distribution $P_{12}$ when $P_1$ and $P_2$ are macroscopically different. To quantify the similarity between distributions, we calculate the Kullback-Leibler divergence between them. For a distribution $P$ and $Q$, this is defined as
\begin{equation}
  D_{KL}(P||Q) = \sum_{\mathbf{x}} P(\mathbf{x}) \log \left( \frac{P(\mathbf{x})}{Q(\mathbf{x})} \right).
\end{equation}
Note that this metric is not symmetric under exchange of $P$ and $Q$. This divergence is the average difference between the log-likelihoods of $P$ and $Q$ under the distribution $P$. In Fig.~\ref{fig:histogram} we show the distributions of $\log \left( P_1 / P_2 \right)$ and $\log \left( P_2 / P_1 \right)$, for samples taken from $P_1$ and $P_2$ respectively. Indeed, the probabilities of samples taken from one distribution are on average 10 orders of magnitude higher than the probabilities according to the other distribution. This confirms that $P_1$ and $P_2$ are highly dissimilar. We also show the distribution of $\log \left( P_{12} / (P_1/2 + P_2/2) \right)$. This distribution is highest at zero, i.e. where both probabilities are equal. The mixture model $P_1/2 + P_2/2$ thus captures correctly the distribution containing both modes $P_{12}$. Note that the difference in number of parameters is significant, i.e. the mixture of two $D=2$ PEPSs has 8 times less parameters than the single $D=4$ PEPS.

\begin{figure}
\includegraphics{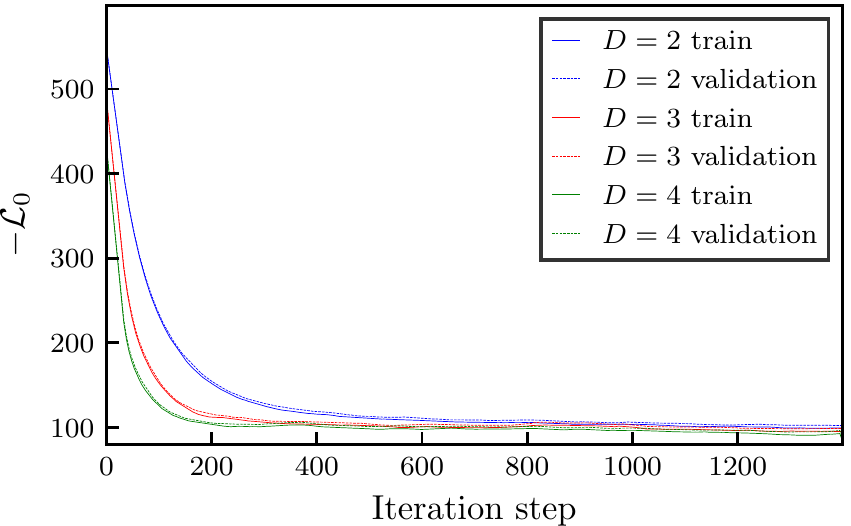}
\caption{Negative log-likelihood as a function of iteration steps on one mode of the binary MNIST dataset. The cost on the training set is indicated with solid lines, that on the validation set with dotted lines.}
 \label{fig:optimize}
\end{figure}

\begin{figure}
\includegraphics{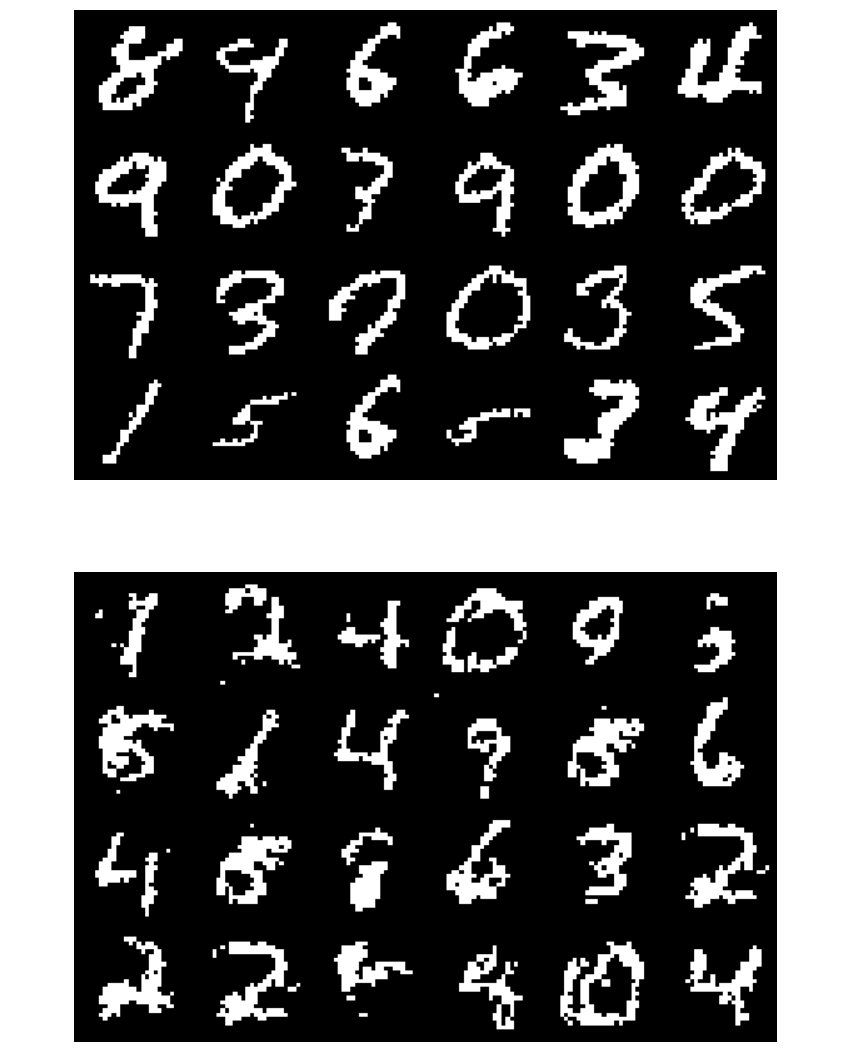}
\caption{Upper panel: images sampled randomly from the training set. Lower panel: Images sampled from the optimized PEPS mixture with $D=4$.}
 \label{fig:digits}
\end{figure}

Next, we proceed with optimizing the likelihood in every mode of the dataset. For details about the optimization procedure, see Appendix. In Fig.~\ref{fig:optimize} we show an example of the decay of the negative log-likelihood of data in one of the modes as a function of iteration steps, for different bond dimensions of the PEPS. The negative log-likelihood improves systematically with increasing bond dimension. We also observe that models with larger bond dimensions converge more rapidly to a fixed value of the cost function. An interesting observation is that the negative log-likelihood of the training set and that of the validation set stay close together during the full optimization procedure. This shows that the model is not overfitting the dataset, even when using the largest bond dimension. Consequently, we can still gain accuracy by increasing the bond dimension further, although this becomes computationally harder very quickly because of the scaling of PEPS algorithms. The absence of overfitting can be contrasted with other tensor-network methods such as MPS, where a significant level of overfitting was observed. This can be attributed to the fact that in the case of MPS, a relatively large numbers of parameters are needed to fit the data. Combined with the fact that an MPS can capture a dataset by remembering every data sample when $D>N_\textrm{data}$, the MPS is indeed prone to overfitting. 

\begin{center}
\begin{table}
\begin{tabular}{ | l | c | } 
 \hline
 model & negative log-likelihood \\ 
 \hline
 MPS \cite{Han2018} & 101.5 \\
 1D TTN \cite{Cheng2019} & 96.9 \\
 2D TTN \cite{Cheng2019} & 94.3 \\ 
 PEPS ($D=2$) & 97.4 \\ 
 PEPS ($D=3$) & 93.2 \\
 PEPS ($D=4$) & 91.2 \\
 \hline
\end{tabular}
\caption{Negative log-likelihoods on the test set of the binarized MNIST dataset for different model architectures.}
\label{tab:nll}
\end{table}
\end{center}

Even with small bond dimensions, our model based on PEPS is able to reach a better log-likelihood than other tensor-network-based algorithms. In Tab.~\ref{tab:nll} we compare the negative log-likelihood we obtain with those other models. Even though the scaling of the number of parameters in PEPS is higher than other tensor network algorithms, the bond dimensions needed are a lot smaller. For example, the MPS results in Tab.~\ref{tab:nll} were obtained with $D_{max}=100$, leading to 20,000 parameters per site. The TTN results, with $D_{max}=50$, lead to 125,000 parameters per tensor. For our PEPS, with $D=4$, we get 512 parameters per site per mode. Even using 10 modes, we only use 5,120 parameters per site.
\par Finally, in order to show the generative power of our PEPS, in Fig.~\ref{fig:digits}, we show examples sampled from our optimized PEPS distributions. For reference, we also show a number of training examples.

\par\noindent\emph{\textbf{Discussion and Conclusion---}} %
%
In this paper, we have presented a two-dimensional tensor network approach to generative modeling of image datasets. The two-dimensional nature of PEPS allows to efficiently capture the horizontal and vertical correlations in datatsets, and treats them on equal footing. Furthermore, the structure of the dataset allows to make the PEPS construction more efficient in utilizing its variational freedom by identifying modes in the data.

Our model is able to generalize well, especially compared to intrinsic one-dimensional tensor networks applied to two-dimensional data. Our approach also reaches larger likelihoods, even with less variational freedom, pointing to a more efficient and more accurate parameterization of the probability distributions. Still, a gap remains compared to the state of the art methods on this problem, including PixelCNN \cite{jain20b}, PixelRNN \cite{pmlr-v48-oord16}, neural autoregressive density estimators \cite{Raiko14}, and also physically-inspired models such as RBMs \cite{Salakhutdinov2008}. This may point in the direction that correlations beyond area law are needed to capture the dataset more accurately \cite{lu2021tensor, martyn2020entanglement}. Another possibility is that algorithms tailored to this problem are needed to find the variational optimum of the PEPS tensor networks. Recently, a generalization of the tensor-network formulation of RBMs has been used to approach the state of the art \cite{liu2021tensor}. This method used a series of tensor networks as its base to capture correlations, and the probabilistic distribution consists of an aggregate of these networks, which is beyond ''pure'' tensor network capabilities. Remarkably, the tensor networks were one-dimensional in nature. This may point in the direction that a suitable combination of pure two-dimensional tensor networks with an appropriate post-processing to construct the actual probability distribution may prove to be the most efficient way to capture probability distributions encountered in generative machine learning.

\par Our observation that a tensor network describing a multimodal distribution decomposes into a sum of tensor networks with smaller bond dimension by bringing them in a block-diagonal form might be used as an unsupervised way to perform clustering of data based on multimodality of the underlying distribution. This requires an algorithm to fix the gauge degrees of freedom of the tensor network such that the tensors take on a block-diagonal form. We will pursue this idea in future work.

\begin{acknowledgements}
This work has received funding from the European Research Council (ERC) under the European Unions Horizon 2020 research and innovation programme (grant agreement No 647905 (QUTE)), and from Research Foundation Flanders (FWO) via grants FWO18/ASP/279 and FWO20/PDS/115. Computational resources (Stevin Supercomputer Infrastructure) and services used in this work were provided by the VSC (Flemish Supercomputer Center), and the Flemish Government – department EWI.
\end{acknowledgements}

\bibliography{ref.bib}

\revappendix
\section{Data set and preprocessing}
The MNIST dataset consists of images of handwritten digits with $28 \times 28$ pixels.  Every pixel has a greyscale intensity, i.e. it has a value between 0 and 255, where 0 is completely black and 255 completely white.  For every digit, the dataset contains 6000 images in the training set and 1000 images in the test set.  Because the local degrees of freedom in our PEPS model take the structure of a $d$-dimensional vector space, we choose to binarize this dataset such that the pixel values are either $0$ or $1$.  Specifically, we choose an often-used binarization as our dataset \cite{Salakhutdinov2008}, where the pixel values were set to one with a probability proportional to its greyscale intensity.  Furthermore, this dataset is unlabeled, and partitioned in a training set of 50000 examples, a validation set of $10000$ examples and a test set of $10000$ examples.  

Because the dataset is unlabeled, we use the second approach described above to identify which example belongs to which mode.  Specifically, we use the UMAP algorithm to find an embedding of the original data points in a two-dimensional space such that they separate in disjunct clusters.  When an embedding is found, we use the DBSCAN algorithm to label every data point with its corresponding cluster.  The DBSCAN algorithm also assigns a label to noisy data points, i.e. those that appear to not be part of a cluster.  We chose to assign these data points to the cluster of its closest neighbor in the two-dimensional space.  Note that both the UMAP and the DBSCAN algorithm have hyperparameters that change the obtained clustering.  In our case, we tuned the hyperparameters to end up with $10$ distinct clusters, as can be expected for the MNIST algorithm.  We manually observed that this clustering coincides with one cluster per digit.  Note that this does not mean that one needs to know the number of modes in advance; taking the number of modes in our model equal to the true number of modes in the data distribution provides the most efficient model in terms of number of parameters, but the model is still capable of approximating the distribution when another partitioning into modes is used.

\section{Optimization details}
We use the following optimization routine.  First, we initialize the PEPS with random weights from a uniform distribution between 0 and 1.  Then, we normalize the probability distribution, which can be performed stochastically by sampling a number of independent configurations to estimate the norm of the PEPS, and then rescaling the weights such that the norm would be equal to one.  Then, we optimize the weights of the PEPS tensor network by calculating the gradient with a batch size of $200$ examples for both the positive phase and negative phase.  We use this gradient to calculate a direction according to the adam algorithm \cite{kingma2014adam}.  The learning rate we chose is $0.001$ and momentum factors were $\beta_1=0.9$ and $\beta_2=0.999$.  When the log-likelihood converges (typically after less than $1000$ iterations), we optimize further with a batch size of $1000$ for the negative phase of the gradient.  Note that two sources of stochasticity appear in the gradient: the effect of taking a batch of examples rather than the full data set and the number of samples used to estimate the gradient of the norm.  Upon increasing the batch size of the negative phase, we reduce the latter type of noise.

\end{document}